\documentclass[a4paper,12pt]{article}

\usepackage[center]{titlesec}

\renewcommand{\thesection}{\Roman{section}}

\renewcommand{\thesubsection}{\arabic{subsection}}

\renewcommand{\thesubsubsection}{\alph{subsubsection}}

\titleformat%
    {\section}
    [block]
    {\centering\bfseries}
    {\thesection.}
    {0.5em}
    {}

\titleformat%
    {\subsection}
    [block]
    {\centering\itshape}
    {\thesubsection.}
    {0.5em}
    {}

\titleformat%
    {\subsubsection}
    [block]
    {\centering}
    {\thesubsubsection.)}
    {0.5em}
    {}

\usepackage{lipsum}

\titlespacing{\section}{0pt}{*4}{*1.5}
\titlespacing{\subsection}{0pt}{*4}{*1.5}
\titlespacing{\subsubsection}{0pt}{*4}{*1.5}

\usepackage[utf8]{inputenc}
\usepackage{cancel}
\usepackage{tikz}
\usepackage{ulem}
\usepackage{amsfonts}
\usepackage{amssymb}
\usepackage{graphicx}
\usepackage{amsmath}
\usepackage{enumerate}
\usepackage{mathtools}
\usepackage{subfig}
\usepackage{color}
\usepackage{tikz}
\usepackage{float}
\usepackage{here}
\usepackage{cite}
\usepackage{mathrsfs}
\usepackage{float,epsfig}
\usepackage{dcolumn}
\usepackage{graphicx}
\usepackage{bm}
\usepackage{amsmath,amssymb,amsthm}
\usepackage[colorlinks=true,linkcolor=blue,citecolor=red]{hyperref}
\usepackage{multirow}
\usepackage[toc,page]{appendix}
\usetikzlibrary{arrows.meta}
\usetikzlibrary{bending}
\usetikzlibrary{calc}
\newcommand{\be}{\begin{equation}}
\newcommand{\ee}{\end{equation}}
\newcommand{\bea}{\setlength\arraycolsep{2pt} \begin{eqnarray}}
\newcommand{\eea}{\end{eqnarray}}

\def\0{{\sst{(0)}}}
\def\1{{\sst{(1)}}}
\def\2{{\sst{(2)}}}
\def\3{{\sst{(3)}}}
\def\4{{\sst{(4)}}}
\def\5{{\sst{(5)}}}
\def\6{{\sst{(6)}}}
\def\7{{\sst{(7)}}}
\def\8{{\sst{(8)}}}
\def\sst#1{{\scriptscriptstyle #1}}

\makeatletter \@addtoreset{equation}{section}

\definecolor{lime}{HTML}{A6CE39}

\begin{document}

\title{{\normalsize \textbf{\Large Dunkl-Corrected Deformation  of  RN-AdS Black Hole Thermodynamics }}}
\author{ {\small  Maryem Jemri\footnote{maryem.jemri@um5r.ac.ma} \hspace*{-8pt}} \\
{\small  ESMaR, Faculty of Science, Mohammed V University in Rabat, Rabat, Morocco}}
\maketitle

\begin{abstract}
In this work,  we derive a new class of  charged black holes by introducing Dunkl derivatives in the  four dimensional spacetime. To construct such solutions, we first compute the Ricci tensor and  the Ricci scalar using the Christoffel symbols.  Substituting  them into the modified Einstein field equations via  extended Dunkl derivations, we obtain the metric function  of  charged Dunkl   black holes.    Next,   we  investigate the charge effect on the   corresponding  thermodynamical properties by computing the    associated quantities. To study  the  thermal stability, we calculate  the heat capacity.  After that, we  approach   the $P$-$v$ criticality behaviors  by  determining  the critical pressure
$P_c$,  the critical temperature $T_c$ and the  critical specific volume $v_c$ in terms of  $Q$ and  two parameters $A$ and $B$ carrying data on the  Dunkl reflections. Precisely, we  show that  the ratio $ \dfrac{P_{c}v_{c}}{T_{c}}$ is a universal number with respect to the charge $Q$  and $B$ parameters.  Taking  a zero limit of  $A$,  we recover    the Van der Waals fluid    behaviors.  For  Joule-Thomson expansion effects  for  such charged  black holes, we reveal  certain   similarities and  the differences with  Van der Waals fluids. Finally,  we discuss the  phase transitions via the Gibbs free energy computations.
{\noindent} 

\end{abstract}

%

\newpage


\newpage

\section{Introduction}
   Black holes are exact solutions of Einstein's field equations within the context of general relativity \cite{1,102}. Their existence has been verified by visionary empirical  observations. Cygnus X-1, which is  a compact object with mass greater than the Tolman-Oppenheimer-Volkoff limit for neutron stars, was the first stellar-mass black hole to be definitively detected with the Isaac Newton Telescope \textbf{\cite{2}}. The dynamical impact of black hole mergers on  the spacetime was further confirmed decades later when the Laser Interferometer Gravitational-Wave Observatory opened a new window by directly observing gravitational waves from these mergers \cite{3}. The first-ever image of a black hole's shadow was recently taken by the Event Horizon Telescope (EHT)\cite{4}, offering a direct visual confirmation of the event horizon. This has been seen as supporting fundamental general relativity predictions in the strong field regime. The implications of these findings for the spacetime geometry and as possible probes of alternative theories of gravity have sparked a flurry of research in black hole physics.

Beside optical  behaviors\cite {74,75,76,77, 077,78},  black holes have been viewed as thermodynamic systems in physical studies \cite{6,71,72,73,79,80,81,82,83}, which means that they have the same laws as classical thermodynamic systems.
Hawking's theoretical prediction of black hole radiations, predicated on further work incorporating quantum effects, shows black holes to possess a temperature that is proportional to their surface gravity \cite{7} and an entropy \cite{8} that depends on it. These developments revolutionized  knowledge of  the spacetime, the laws of physics, and the basics that govern the universe. In particular, the electric charge is found to have a remarkable impact on the black holes thermodynamical characteristics such as temperature, entropy, and heat capacity, while also leading to intricate phase structures and critical phenomena.

Moreover, a  special emphasis has been placed on investigating black hole properties arising from non-trivial spacetime geometries by introducing additional deformation parameters. In this context, black hole solutions have been constructed within a gauge theory framework for gravity, where the de Sitter group in four dimensions is treated as a local gauge symmetry \cite{888}. By coupling gravity with spacetime deformation effects, various black hole configurations have been obtained. In particular, black holes in non-commutative spacetimes have been studied by incorporating relevant geometric quantities as central elements in gravitational computations based on the Einstein field equations \cite{889}.

Recently, a study introduced a deformed Schwarzschild black hole within the framework of de Sitter gauge gravity by incorporating Dunkl-type generalized derivatives to solve the Einstein field equations and derive the corresponding solution \cite{8890,890,891,892}. This work first examined the black hole's thermodynamical characteristics, but it was later expanded to examine other phenomena like phase transitions and black hole shadows \cite{891,892}. Moreover, the Dunkl deformation parameter has been constrained using observational data on supermassive black holes from the EHT data.

The aim of this paper is to contribute to such activities by    providing   a new class of  charged black holes by introducing Dunkl derivatives in the  four dimensional spacetime.  To  obtain  such solutions, we first compute the Ricci tensor and the  Ricci scalar using the Christoffel symbols.  Substituting  them into the modified Einstein field equations  via Dunkl derivatives, we obtain the metric function of  the associated charged  black holes. Then,   we   study the charge effects on the   thermodynamical properties by computing the    associated quantities. To  investigate the  thermal stability, we calculate and examine   the heat capacity variation in terms of the charge $Q$.  After that, we   examine    the $P$-$v$ criticality behaviors  by  determining  the critical pressure
$P_c$,  the critical temperature $T_c$ and the  critical specific volume $v_c$ in terms of  $Q$ and  two parameters $A$ and $B$ carrying data on the  Dunkl reflections.  Concretely, we  reveal  that  the ratio $ \dfrac{P_{c}v_{c}}{T_{c}}$ is an universal number with respect to the charge $Q$  and $B$ parameters.  Taking  a zero limit of  $A$,  we recover    the Van der Waals fluid    behaviors.  For  Joule-Thomson expansion effects  for  such black holes, we  show   certain   similarities and  the differences with  Van der Waals fluids. Finally,  we study the  phase transitions via the Gibbs free energy computations.

The  organization of this work  is as follows.  In section 2,   we present a  new class of deformed charged  AdS black holes from Dunkl  formalisms.
In section 3, we calculate  certain  thermodynamical quantities in order to   investigate  the
stability behaviors.  In  section 4, we   study   the critical aspect by focusing on 
the  $P-v$   diagrams,  the phase transitions, and   the  Joule-Thomson expansion effects. We end  this work by  certain concluding  remarks.

\section{Deformed Charged Black Holes in the Presence of Dunkl Operators}

\begin{figure}[h!]
    \centering
    \begin{tabular}{cc}
       \includegraphics[width=7.5cm,height=7.5cm]{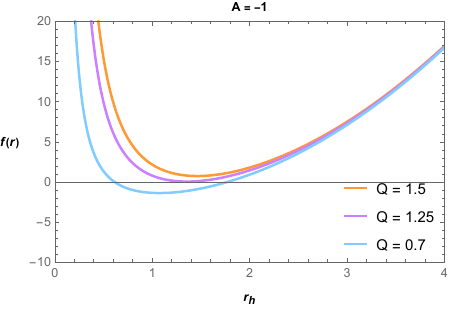} & 
         \includegraphics[width=7.5cm,height=7.5cm]{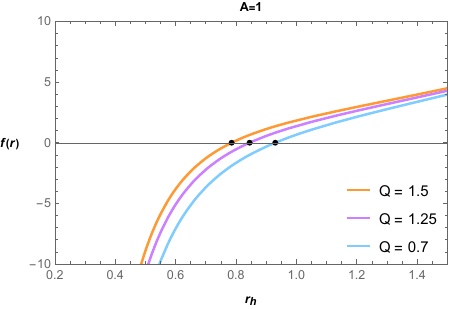} 
    \end{tabular}
     \caption{Effect of the charge parameter $Q$ and the Dunkl parameter $A$ on the metric function $f(r)$ by considering $B=1$.}
    \label{Fig3.1}
\end{figure}

In this section, we aim to construct a novel class of  charged black hole solutions by incorporating Dunkl-type differential operators into the Einstein–Maxwell framework. These operators introduce reflection symmetries into the geometry, thereby modifying the standard structure of the  spacetime. Specifically, we investigate a static, spherically symmetric, charged spacetime and analyze how the presence of Dunkl deformations alters the gravitational and electromagnetic fields. To start,  we consider the following  general metric form 
\begin{equation}
\mathrm{d}s^2 = g_{\mu\nu} \mathrm{d}x^{\mu} \mathrm{d}x^{\nu},
\end{equation}
where $\mathrm{d}s^2$
represents the spacetime interval between two nearby events. $ \mathrm{d}x^{\mu}$ and $\mathrm{d}x^{\nu}$
indicate infinitesimal displacements in each coordinate direction, where the value of indices denote 
the spacetime dimensions. The elements $g_{\mu\nu}$ are the components of the metric tensor, which
can vary from point to point in  the curved spacetimes \cite{893}.  To get  the black hole solutions that we are after,  we focus on  a spherically symmetric metric via the  following  ansatz 
\begin{equation}
\mathrm{d}s^2 =-f(r)\,\mathrm{d}t^2 + \frac{1}{f(r)}\,\mathrm{d}r^2 + r^2\left(\mathrm{d}\theta^2 + \sin^2\theta \,\mathrm{d}\phi^2\right),
\end{equation}
where \( f(r) \) is an unknown radial  function to be determined by solving the Einstein field equations in the presence of the modified geometry induced by Dunkl operators given by
\begin{eqnarray}\label{4}
	D_{x_{i}}= \dfrac{\partial}{\partial x_{i}}+\dfrac{\alpha_{i}}{x_{i}}(1-\mathcal{R}_{i})\quad,\quad(i=0,1,2,3)
	\end{eqnarray}	
where, $\alpha_{i} = (0, \alpha_{1}, \alpha_{2}, \alpha_{3})$ are the Dunkl parameters, constrained by $\alpha_i > -1/2$, and $\mathcal{R}_{i} = (0, \mathcal{R}_{1}, \mathcal{R}_{2}, \mathcal{R}_{3})$ are the associated parity operators, which act as $\mathcal{R} = +1$ for even functions and $\mathcal{R} = -1$ for odd functions \cite{894,895,896,897,898}. The Dunkl operator formalism systematically incorporates these discrete parity transformations and finite reflection symmetries into the study of spacetime geometries. This extends  the framework of differential calculus to settings with reflection group symmetries.	
According to \cite{890}, the components of the Dunkl operators in  the spherical coordinates can be written as 
	\begin{equation}\label{6}
	\begin{aligned}
	&D_{r}=\frac{\partial}{\partial r}+\frac{1}{r}\sum_{i=1}^{3}\alpha_{i}(1-\mathcal{R}_{i}),\quad\quad D_{t}=\frac{\partial}{\partial t},\\
	&D_{\theta}=\frac{\partial}{\partial \theta}+\sum_{i=1}^{2}\alpha_{i}(1-\mathcal{R}_{i}) \cot\theta-\alpha_{3}(1+\mathcal{R}_{3})\tan\theta,\\
	&D_{ \phi}=\frac{\partial}{\partial \phi}-\alpha_{1} \tan\phi (1-\mathcal{R}_{1})+\alpha_{2} \cot\phi (1-\mathcal{R}_{2}).
	\end{aligned}
	\end{equation}
To proceed, we compute the Christoffel symbols associated with the above metric. These symbols, which  play a central role in defining the curvature of  the spacetime, are given by
\begin{equation}
\Gamma^\lambda_{\mu\nu} = \frac{1}{2} g^{\lambda\rho} \left( D_\mu g_{\nu\rho} + D_\nu g_{\mu\rho} - D_\rho g_{\mu\nu} \right).
\end{equation}
However, due to the Dunkl deformation, the standard derivatives are supplemented by reflection terms characterized by parameters \( \alpha_i \) and reflection operators \( \mathcal{R}_i \). As a result, the connection acquires additional contributions that encode the discrete symmetry of the deformation.  After computations, we find that the non-zero modified Christoffel symbols are given by
{\footnotesize
\begin{equation}
\begin{aligned}
\Gamma^{r}_{rr} &= \frac{1}{2} \left( -\frac{f'}{f} + \frac{1}{r} \sum_{i=1}^{3} \alpha_i (1 - \mathcal{R}_i) \right), \\
\Gamma^{r}_{\theta\theta} &= -\frac{f r}{2} \left( 2 + \sum_{i=1}^{3} \alpha_i (1 - \mathcal{R}_i) \right), \\
\Gamma^{\phi}_{\theta\phi} &= \Gamma^{\phi}_{\phi\theta} = -\frac{1}{2} \left( 2 \cot\theta + \delta \right), \\
\Gamma^{r}_{tt} &= \frac{f}{2} \left( f' + \frac{f}{r} \sum_{i=1}^{3} \alpha_i (1 - \mathcal{R}_i) \right), \\
\Gamma^{\theta}_{r\theta} &= \Gamma^{\theta}_{\theta r} = \frac{1}{2r} \left( 2 + \sum_{i=1}^{3} \alpha_i (1 - \mathcal{R}_i) \right), \\
\Gamma^{\theta}_{\phi\phi} &= -\frac{1}{2} \sin^2\theta \left( 2 \cot\theta + \delta \right), \\
\Gamma^{t}_{tr} &= \frac{1}{2} \left( \frac{f'}{f} + \frac{1}{r} \sum_{i=1}^{3} \alpha_i (1 - \mathcal{R}_i) \right), \\
\Gamma^{\phi}_{r\phi} &= \Gamma^{\phi}_{\phi r} = \frac{1}{2r} \left( 2 + \sum_{i=1}^{3} \alpha_i (1 - \mathcal{R}_i) \right), \\
\Gamma^{r}_{\phi\phi} &= -\frac{f r \sin^2\theta}{2} \left( 2 + \sum_{i=1}^{3} \alpha_i (1 - \mathcal{R}_i) \right)
\end{aligned}
\end{equation}
}
where  \( \delta\) is  a parameter, which will be called  Dunkl parameter,   carrying   the effect of the deformation on the spherical symmetry given 
\begin{equation}
\delta= \sum_{i=1}^{2} \alpha_i (1 - \mathcal{R}_i) \cot\theta - \alpha_3 (1 + \mathcal{R}_3) \tan\theta.
\end{equation}
Using the modified Christoffel symbols, we compute the Ricci tensor, which captures the spacetime  local curvature 
\begin{equation}
R_{\mu\nu} = D_\alpha \Gamma^{\alpha}_{\mu\nu} - D_\nu \Gamma^{\alpha}_{\mu\alpha} + \Gamma^{\alpha}_{\mu\nu} \Gamma^{\beta}_{\alpha\beta} - \Gamma^{\alpha}_{\mu\beta} \Gamma^{\beta}_{\alpha\nu}.
\end{equation}
After straightforward but lengthy computations, we obtain the non-zero components which are given by 
\begin{equation}
\begin{aligned}
R_{tt} &= \frac{1}{2} f f'' + \frac{f f'}{r} \left( 1 + \frac{3}{2} \sum_{i=1}^{3} \alpha_i (1 - \mathcal{R}_i) \right) + \frac{f^2}{r^2} \sum_{i=1}^{3} \alpha_i (1 - \mathcal{R}_i) \left( \frac{1}{2} + \sum_{i=1}^{3} \alpha_i (1 - \mathcal{R}_i) \right), \\
R_{rr} &= -\frac{1}{2} \frac{f''}{f} - \frac{f'}{f r} \left( 1 + \frac{3}{2} \sum_{i=1}^{3} \alpha_i (1 - \mathcal{R}_i) \right) - \frac{3}{2 r^2} \sum_{i=1}^{3} \alpha_i (1 - \mathcal{R}_i) \left( 1 + \sum_{i=1}^{3} \alpha_i (1 - \mathcal{R}_i) \right), \\
R_{\theta\theta} &= -f - r f' + 1 - \frac{r f'}{2} \sum_{i=1}^{3} \alpha_i (1 - \mathcal{R}_i) - \frac{5}{2} f \sum_{i=1}^{3} \alpha_i (1 - \mathcal{R}_i) - f \left( \sum_{i=1}^{3} \alpha_i (1 - \mathcal{R}_i) \right)^2 + B, \\
R_{\phi\phi} &= R_{\theta\theta} \sin^2\theta,
\end{aligned}\label{ii}
\end{equation}
where $B$ is  a  angular correction parameter  expressed as 
\begin{equation}
B =\delta \left( \frac{\delta}{2} + 2 \cot\theta \right).
\end{equation}
It is denoted that the Ricci scalar, obtained by contracting the Ricci tensor with the inverse metric, takes the form
\begin{equation}
R = g^{tt} R_{tt} + g^{rr} R_{rr} + g^{\theta\theta} R_{\theta\theta} + g^{\phi\phi} R_{\phi\phi}.
\end{equation}
The computations provide 
\begin{equation}
R =-f'' - \frac{2f}{r^2} + \frac{2(1 + B)}{r^2} - \frac{4f'}{r} \left( 1 + \sum_{i=1}^{3} \alpha_i (1 - \mathcal{R}_i) \right)
- \frac{f}{r^2} \sum_{i=1}^{3} \alpha_i (1 - \mathcal{R}_i) \left( 7 + \frac{9}{2} \sum_{i=1}^{3} \alpha_i (1 - \mathcal{R}_i) \right).
\end{equation}
To unveil the charge effect, we need to  introduce  the  electromagnetic field coupled to the  gravity.  Following  \cite{219}, this dynamics  is described by the Einstein-Maxwell action
 \begin{equation}
  S = \frac{1}{16\pi G} \int d^4 x \sqrt{-g} \left( R - F^{\mu\nu} F_{\mu\nu} \right)
\end{equation} 
 where $F^{\mu\nu}$ is Maxwell’s electromagnetic field tensor, related to the electromagnetic potential
 ${\cal A}_{\mu}$ by the following equation
 \begin{equation}
  F_{\mu\nu} = D_{\mu} {\cal A}_{\nu} - D_{\nu} {\cal A}_{\mu}.\label{1} 
\end{equation}  
  By considering spherical symmetry, the electromagnetic potential $A_{\mu}$ is given by
  \begin{equation}
  {\cal A}_\mu = ({\cal A}_t(r), 0, 0, 0) 
\end{equation}    
  where one has used 
    \begin{equation}
      {\cal A}_t(r) = \frac{Q}{4\pi\varepsilon_0 r}.
\end{equation}    
  Varying  $S$  with respect to the metric $g_{\mu \nu}$ gives the energy-momentum tensor in terms of the 
 Maxwell’s electromagnetic field tensor
 \begin{equation}
  T_{\mu\nu} = \frac{1}{4\pi} \left( F_{\rho\mu} F^{\rho\beta} g_{\nu\beta} - \frac{1}{4} g_{\mu\nu} F_{\rho\beta} F^{\rho\beta} \right).\label{2}
\end{equation} 
 From the Eq.(\ref{1}),  we find the components of $F_{\mu\nu}$
 \begin{equation}
 \begin{aligned}
F_{tr} &= -D_r {\cal A}_t(r) = \frac{Q}{4\pi\varepsilon_0 r^2}\left(1-\sum_{i=1}^{3}\alpha_{i}(1-\mathcal{R}_{i})\right)^{\!2} \\
F_{rt} &= -F_{tr} = -\frac{Q}{4\pi\varepsilon_0 r^2}\left(1-\sum_{i=1}^{3}\alpha_{i}(1-\mathcal{R}_{i})\right)^{\!2}.
\end{aligned}
 \end{equation}
Taking the  contra-variant components of $
 F^{\mu\nu} = g^{\mu\alpha} g^{\nu\beta} F_{\alpha\beta}$, 
we get 
\begin{equation}
F^{rt} = -F^{tr} = \frac{Q}{4\pi\varepsilon_0 r^2}\left(1-\sum_{i=1}^{3}\alpha_{i}(1-\mathcal{R}_{i})\right)^{\!2}.
\end{equation}
From the Eq (\ref{2}),  we find that the non-zero components of $T_{\mu \nu}$ are given by 
 \begin{equation}
  \begin{aligned}
T_{tt} &= \frac{Q^2 f}{32\pi^3\varepsilon_0^2r^4}\left(1-\sum_{i=1}^{3}\alpha_{i}(1-\mathcal{R}_{i})\right)^{\!2} \\
T_{rr} &= \frac{-Q^2}{32\pi^3\varepsilon_0^2r^4 f}\left(1-\sum_{i=1}^{3}\alpha_{i}(1-\mathcal{R}_{i})\right)^{\!2} \\
T_{\theta\theta} &= \frac{Q^2}{32\pi^3\varepsilon_0^2r^2}\left(1-\sum_{i=1}^{3}\alpha_{i}(1-\mathcal{R}_{i})\right)^{\!2} \\
T_{\phi\phi} &= \sin^2\theta\, T_{\theta\theta}.
\end{aligned}\label{iii}
 \end{equation}
Substituting  the curvature components and the energy-momentum tensor into Einstein’s field equations
\begin{equation}
G_{\mu\nu} = R_{\mu\nu} - \frac{1}{2} R g_{\mu\nu} = 8\pi G T_{\mu\nu}, \label{ei}
\end{equation}
we derive the  following set of modified differential equations for the black hole metric function  \( f(r) \)
\begin{align}
-\frac{f'}{2r}(2 + A) - \frac{3f}{r^2}A - \frac{5f}{4r^2}A^2 - \frac{f}{r^2} + \frac{1 + B}{r^2} &= \frac{Q^2}{r^4}(1 - A)^2, \label{11}\\
-\frac{f' r}{2}(2 + 3A) + f A \left(1 + \frac{5}{4}A\right) + \frac{r^2 f''}{2} &= \frac{Q^2}{r^2}(1 - A)^2,\label{22} 
\end{align}
where one has  used a new Dunkl parameter  
\begin{equation}
A = \sum_{i=1}^3 \alpha_i(1 - \mathcal{R}_i) 
\end{equation}
 and $\dfrac{G}{4 \pi^{2} \varepsilon^{2}_{0}}=1$. By summing Eqs. (\ref{11}) and (\ref{22}), we obtain the following unified expression
\begin{equation}
r^2 f'' - 2f (1 + 2A) + 2r f' A + 2(1 + B) = \frac{4Q^2}{r^4}(1 - A)^2.
\end{equation}
Solving this system leads to the deformed  charged metric function
\begin{equation}
f(r) = \frac{1 + B}{1 + 2A} + \frac{c_1}{r^{1 + 2A}} +c_2 r^2 + \frac{Q^2}{r^2} \left(1 - \frac{A^2}{1 + 2A} \right),\label{o}
\end{equation}
where $c_1$ and $c_2$ are integration constants.
 Putting   $c_1=-2M\epsilon_M$ and $c_2=- \frac{\Lambda}{3}$, we obtain 
\begin{equation}
f(r) = \frac{1 + B}{1 + 2A} - \frac{2M\epsilon_M}{r^{1 + 2A}} - \frac{\Lambda}{3} r^2 + \frac{Q^2}{r^2} \left(1 - \frac{A^2}{1 + 2A} \right),\label{o}
\end{equation}
where \( M \) and \( \Lambda \) are integration constants interpreted  as the black hole mass and the cosmological constant, respectively.  $\epsilon_M$ is constant with dimension of [$L] ^{2A}$.  For simplicity reason, we consider  $\epsilon_M=1$, the obtained charged  black hole  reduces to
\begin{equation}
f(r) = \frac{1 + B}{1 + 2A} - \frac{2M}{r^{1 + 2A}} - \frac{\Lambda}{3} r^2 + \frac{Q^2}{r^2} \left(1 - \frac{A^2}{1 + 2A} \right),\label{o}
\end{equation}
The resulting solution Eq. (\ref{o}) indeed satisfies the full original system of Einstein equations Eq. (\ref{ei}), together with the conditions Eqs. (\ref{ii}) and (\ref{iii}). Moreover, under the assumed symmetry and metric ansatz, this solution is the unique consistent solution of the system.
 In the absence of  Dunkl deformations,  \( A = 0 \) and \( B = 0 \), the metric  function \( f(r) \) reduces to
\begin{equation}
f(r) = 1 - \frac{2M}{r} - \frac{\Lambda}{3} r^2 + \frac{Q^2}{r^2},
\end{equation}
which corresponds to the well-known Reissner–Nordström–de Sitter solution \cite{220}. This confirms the consistency of the present  deformed geometry in the appropriate limit.
At this level, we would like to provide a comment.  It has been observed that  the limit $Q=0$  does not recover   the solution  elaborated  in \cite{890} given by 
\begin{equation}
f_{\xi}(r)=\frac{1}{(1+\xi)}-2 M r^{\frac{1}{2}(1-\sqrt{9+8 \xi})}-\frac{\Lambda}{3} r^{\frac{1}{2}(1+\sqrt{9+8 \xi})}, \label{d}
\end{equation}
where  $\xi$  is a parameter given in terms of Dunkl reflections. We  expect that  the presence of the electric charge $Q$  has played a significant role in the deformation structure of  the obtained charged solutions. However,  a possible link could be worked  out  by considering  $\Lambda=0$ and $Q=0$, for the two solutions given by 
\begin{equation}
f_{\xi}(r)=\frac{1}{(1+\xi)}-2 M r^{\frac{1}{2}(1-\sqrt{9+8 \xi})}
\end{equation} 
and 
\begin{equation}
f(r) = \frac{1 + B}{1 + 2A} - 2M r^{-(1 + 2A)},
\end{equation} 
 By performing a limiting expansion of the functions
$\frac{1}{(1+\xi)}$ and $\frac{1}{2}(1-\sqrt{9+8 \xi})$
\begin{equation}
\frac{1}{1+\xi} = 1 - \xi + O(\xi^{2}), 
\qquad 
\frac{1}{2}\!\left( 1 - \sqrt{9 + 8\xi} \right)
= -1 - \frac{2}{3}\,\xi + O(\xi^{2}).
\end{equation}
we find that the equivalence is recovered when the deformation parameters $A$ and $B$ satisfy the conditions 
\begin{equation}
A=\dfrac{\xi}{3}, \qquad 
B=-\dfrac{\xi}{3}(2\xi+1).
\end{equation}
 This confirms that the discrepancy arises from the interplay between the electric charge and the Dunkl deformations.\\
Before discussing  the thermodynamic  properties of the obtained charged Dunkl solutions,  we first  examine  the  black hole metric function behaviors.  Fixing the mass and the cosmological constant,  the discussion will be elaborated in terms  of 
three relevant parameters: $(A, B,Q)$.  Roughly,   Fig.(\ref{Fig3.1}) illustrates  such behaviors. 

 It has been observed that,  fixing the  $A$  and  $B$  parameters,  there exists a critical charge value denoted by  $Q_c$  associated with a double zero of $f(r_h)=0$ providing an extremal    black hole. Moreover, they  are two  horizons (the inner and outer horizons) and a  naked singularity.  For  $Q>Q_{c}$,   a naked
singularity appears. For $Q<Q_{c}$, however, one has  a   solution  describing  the 
non-extremal  black holes.  In the rest of this  work, we consider only physical solutions, which we will refer to as  charged Dunkl black holes.

\section{Thermodynamics and Stability of Charged Dunkl  Black Holes}
\begin{figure}[h!]
    \centering
    \begin{tabular}{cc}
       \includegraphics[width=7.5cm,height=7cm]{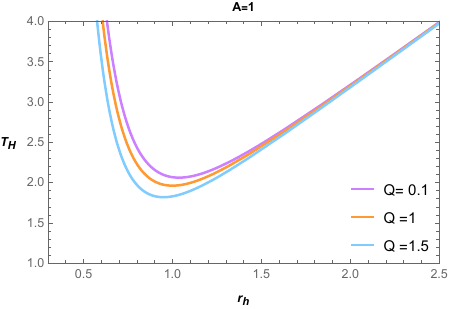} & 
         \includegraphics[width=7.5cm,height=7cm]{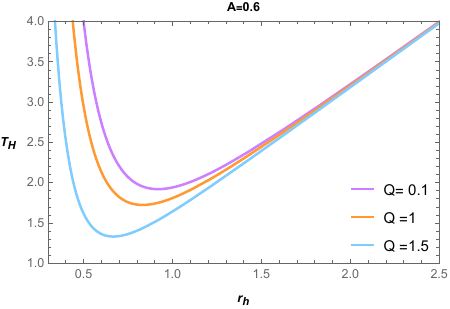} 
    \end{tabular}
    \caption{ Charge parameter effect $Q$  on  the  temperature  by taking $B=1$ for two different values of $A$.}
    \label{Fig3.2}
\end{figure}

In this section, we would like to investigate  the charged Dunkl black holes  by approaching   certain thermodynamic behaviors including the stability aspect. To do so, we first need  to compute  the relevant thermodynamic quantities such  as  the temperature and the heat capacity.  To  find the associated expressions,    one should determine    the mass quantity.  In particular, we find the mass as a function of the horizon radius $ r_h $ taking into account the constraint $ f(r) = 0 $. With regard to the mass, we find that it is given by 
\begin{equation}
M = \frac{ -\Lambda \left(1+2A \right) r_h^4 + 3 \left(B+1 \right) r_h^2 - 3Q^2\left( A^2-2A-1 \right)  }{ 6 \left(1+2A \right) r_h^{1-2A} }.    
\end{equation}
Concerning  the  Hawking temperature, one should  exploit  $T_H = \kappa / (2\pi) $, where the surface gravity $ \kappa $  reads as 
\begin{equation}
\kappa = \left. \frac{ \mathrm{d}f(r) }{ \mathrm{d}r } \right|_{r = r_h}.    
\end{equation}
The computations provide the following  Hawking temperature
\begin{equation}
T_H =\frac{-\Lambda  \left(1+2A\right) \left( 3+ 2A \right) r_{h}^{4}+3 \left(1+2 A \right)\left(1+B \right) r_h^{2}-3 Q^{2} \! \left(2 A^{3}-5 A^{2}+1\right)}{12  \pi  \left(1+2 A \right) \,r_h^{3}}.  \label{xxx} 
\end{equation}
This expression recovers certain known results.  Taking $ A=B=Q= \Lambda=0$,  we find  the Schwarzschild black
hole temperature  being $T_{H}=\frac{1}{4\pi r_h}$ \cite{221}.
To examine the thermal variation, we graph the temperature as a function of the radius of the event horizon in Fig.(\ref{Fig3.2}) by taking different points in the space parameter.

 The plot shows how the charge $ Q $ and the Dunkl parameter $ A $ affect the black hole temperature $ T_H $ as a function of the horizon radius $ r_h $, where one has used  $ B = 1 $.  As  the charge $ Q $ increases, the temperature decreases for small $ r_h $, and the minimum of $ T $ shifts to larger $ r_h $. Comparing the two panels, we observe that when $ A $ decreases from 1 to 0.6, the temperature minimum becomes deeper and the low-temperature region widens. This indicates that both parameters modify the thermal behavior, $ Q $ lowers the temperature, while smaller values of $ A $ enhance this effect and alter the shape of the $ T_H- r_h $ curve.\\
Having discussed the thermal behavior, we move now to inspect  the   local  thermodynamic stability of  the obtained charged Dunkl black holes. Thus, we should calculate the heat capacity  $C_p$   
given by 
\begin{equation}
C_p = T_H  \frac{\partial S}{\partial T_H},
\end{equation}
where one has used the  following entropy 
\begin{equation}
S = \frac{r_h^{2(1+A)} \pi}{1+A}.\label{yyy}
\end{equation}
It is denoted  that in the absence of Dunkl parameter ($A = 0$), and in  the standard gravity,  this entropy reduces to
\begin{equation}
S = \pi r_h^2.
\end{equation}
Using the standard computations, the heat capacity is found to be 
\begin{equation}
C_p = \frac{2 \pi \left(-\Lambda \!  \left(1+2A\right) \left( 3+ 2A \right) r_h^{4}+3 \left(1+B \! \left(1+2 A \right)\right) r_h^{2}-3 Q^{2} \! \left(2 A^{3}-5 A^{2}+1\right)\right) \,r_h^{2(1+A)} }{-\Lambda \! \left(1+2A\right) \left( 3+ 2A \right) r_h^{4}-3 \left(1+B \! \left(1+2 A \right)\right) r_h^{2}+9 Q^{2} \! \left(2 A^{3}-5 A^{2}+1\right)}.
\end{equation}
Considering $ A=B=Q= \Lambda=0$,  we find  the capacity of the Schwarzschild black
hole   given by 
\begin{equation}
C_p =-2  \pi r_h^{2} .
\end{equation}
Based on the sign of the thermal capacity, we can check the stability of the corresponding black hole solutions: a locally stable thermodynamic system can occur if $C_{p}>0$, while an unstable solution occurs if $C_{p}<0$. A graphical representation  is given in Fig.(\ref{4}), where we illustrate \( C_{p} \) as a function of \( r_{h} \) for selected points in the parameter space.
\begin{figure}[h!]
    \centering
    \begin{tabular}{cc}
    \includegraphics[width=8cm,height=7cm]{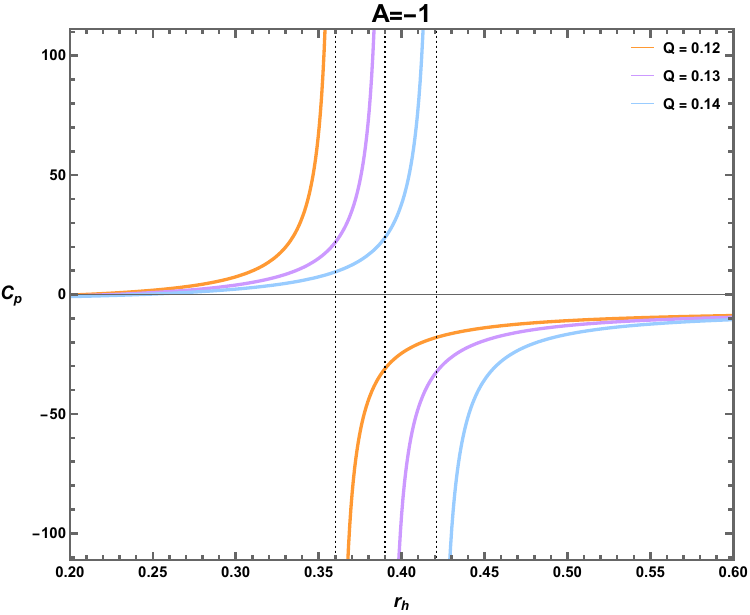} 
    \includegraphics[width=8cm,height=7cm]{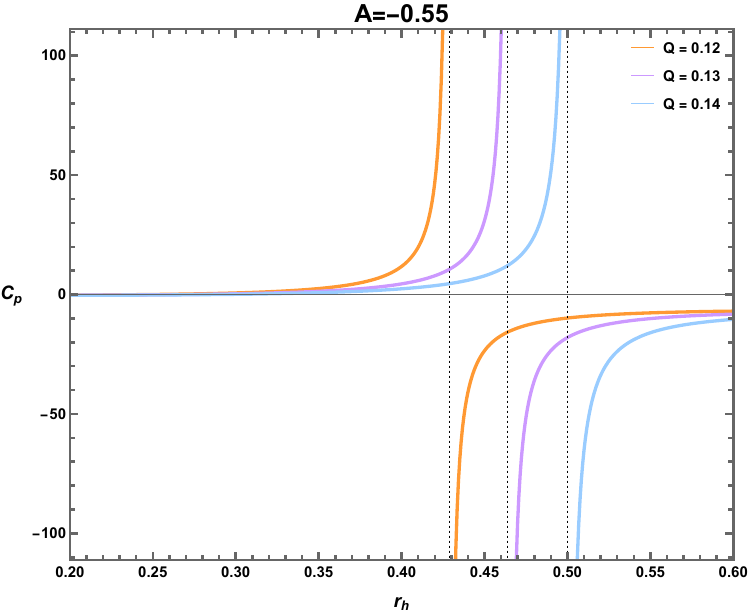} 
    \end{tabular}
    \caption{Effect of the  charge  parameter \( Q \) on the  heat  capacity by taking $B=1$ for two different values of $A$}
    \label{4}
\end{figure}
For a general point in  the parameter space, we observe  that the heat capacity curves are discontinuous at the critical values $ r_{h} = r_{h}^{c} $. At these points, the heat capacity $ C_p $ exhibits divergent behaviors, clearly indicating a second-order phase transition. By setting the deformation parameter $ A = -1 $, we observe that the critical radius $ r_{c} $ increases with the charge $ Q $. Around each divergence point, two branches appear. For $ r_h < r_h^c $, the thermal capacity is negative, indicating the  thermodynamic instability. For $ r_h > r_h^c $, the heat capacity becomes positive, indicating the stable black hole configurations. This confirms that  the charged black holes exhibit both stable and unstable phases, with the phase transition point depending on the value of $ Q $. Furthermore, for a higher deformation parameter, such as $ A = -0.55$, the critical radius $ r_h^c$ is larger for the same value of $Q$. This indicates that as the deformation parameter $ A $ increases, the critical radius $ r_h^c $ decreases.

\section{Critical and Universal Behaviors of Charged Dunkl Black Holes}
In this section, we focus on the critical behaviors,  the Joule-Thompson expansion, and  the phase transitions of these charged  Dunkl black hole solutions by performing calculations of the relevant thermodynamic quantities.
 \subsection{$P$-$v$ criticality behaviors} 
It is essential to determine the thermodynamic state equation of such charged AdS black hole solutions, which can be established by considering the cosmological constant $\Lambda $   as a pressure\begin{equation}
P=-\frac{\Lambda }{8\pi }.
\end{equation}%
After computations, the pressure  is found to be 
\begin{equation}
 P =\frac{3 \left(  4\pi  T  \left(1+2 A \right) r_h^{3}-\left(1+B \right)\left(1+2 A \right) r_h^{2}+ Q^{2} \left(2 A^{3}-5 A^{2}+1\right)\right) }{8\pi   \left(1+2A\right) \left( 3+ 2A \right) r_h^{4}}
.   \label{P}
\end{equation}
Considering  the black hole thermodynamic volume as 
 \begin{equation}
 V=\frac{4 \pi r_h^{3+2 A} }{3},\label{V}
 \end{equation}
we can get   the critical pressure  $P_c$,  the critical specific volume $v_c$
and the  critical temperature $T_c$ by solving the following constraints 
\begin{equation}
\frac{\partial P}{\partial r_h }=0,\hspace{1.5cm}\frac{\partial ^{2}P}{%
\partial r_h^{2}}=0.
\end{equation}
We find that the  critical quantities are  given by 
\begin{equation}
\begin{aligned}
P_c&=\frac{\left(1+2 A \right) \left(B +1\right)^{2}}{32 Q^{2} \pi  \left(3+2 A \right) \left(2 A^{3}-5 A^{2}+1\right)} \nonumber\\ 
T_{c}&=\frac{\sqrt{6}\, \left(B +1\right) \sqrt{\left(B +1\right) \left(1+2 A \right)}}{18 \pi  Q \sqrt{2 A^{3}-5 A^{2}+1}} \nonumber\\
v_{c}&=\frac{2 \sqrt{6}\, Q \sqrt{2 A^{3}-5 A^{2}+1}\,}{\sqrt{(B+1) \! \left(1+2 A \right)}}.
\nonumber
\end{aligned}
\end{equation}
It has been observed that certain constraints must be imposed on the Dunkl parameters $A$ and $B$ in order to obtain real and physically meaningful critical quantities.
In particular,  the real  expressions  require
\begin{equation}
B>1,\qquad 1-\sqrt{2}<A<\dfrac{1}{2},
\end{equation}
which ensure that all square roots are real and that the denominators do not vanish. These restrictions are taken into account in the graphical analysis presented below.
Additional constraints could  also arise from the study of the shadow of Dunkl black holes. In particular,   certain numerical  computations have been performed  using CUDA  code exploited in machine learning methods \cite{BBBJ}. Such an analysis  show that only specific ranges of the Dunkl parameters produce convergent and physically admissible solutions provided by  considering further restrictions  to be taken into account.

The critical triple $(P_{c}, T_{c}, v_{c})$ provides  the following ratio
\begin{equation}
\chi=\dfrac{P_{c}v_{c}}{T_{c}}=\frac{9}{24+16 A},
\end{equation}
which does not keep a fixed value like charged AdS black holes \cite{17}.  However,  this expression  could produce certain known relations. Taking small values of $A$,  we recover   the usual  universal behavior with respect to the electric charge $Q$
\begin{equation}
\chi=\dfrac{3}{8}- \dfrac{1}{4}A+  \dfrac{1}{6}A^2+O( A^3).
\end{equation}
 For   $A = 0$,  we  exactly  obtain   the  RN-AdS black hole situation  \cite{17,177,178}.  
 
\begin{figure}[h!]
    \centering
    \begin{tabular}{cc}
       \includegraphics[width=7.5cm,height=7.5cm]{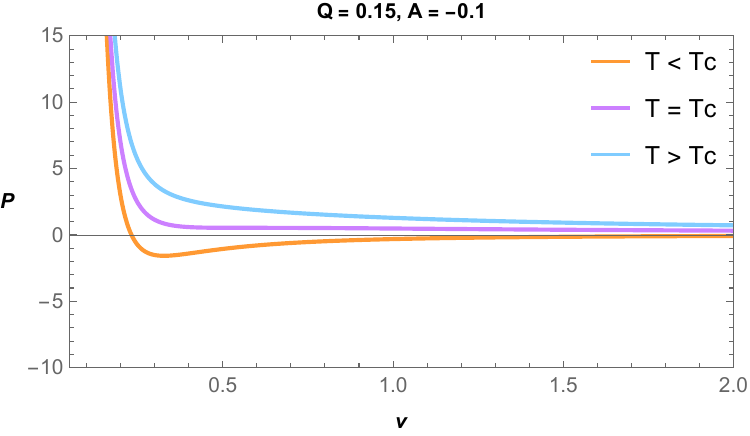} & 
         \includegraphics[width=7.5cm,height=7.5cm]{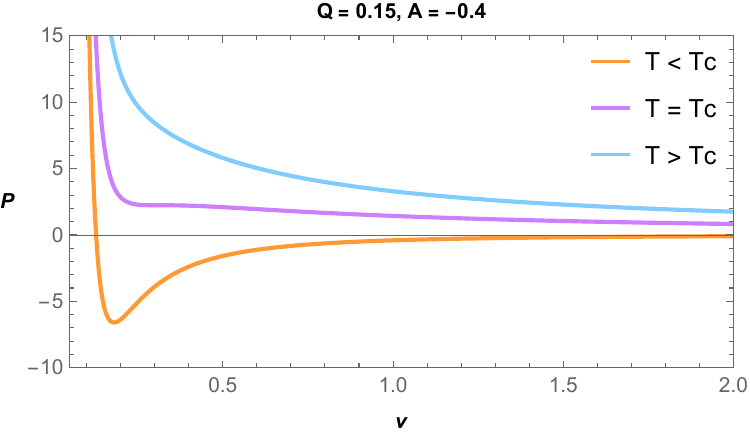} 
    \end{tabular}
\caption{Pressure in terms of $v$ for different values of $T$ and $A$ with $B=1$. }
\label{F41}
\end{figure}

In Fig.(\ref{F41}), we plot the $P-v$ diagram. 
  It is clear  that for a temperature $T$ larger than the critical one $T_c$, the system behaves like an ideal gas. The critical isotherm at $T = T_c$ is characterized by an inflection point at the critical pressure $P_c$ and the critical volume $v_c$. For $T < T_c$, there exists an unstable thermodynamic region. Clearly, the $P - v$ diagram resembles that of a Van der Waals fluid. In addition, we observe that the Dunkl deformation parameter $A$ affects the thermodynamic behavior of the system. As $A$ increases, the minimum value of the pressure $P$ also increases for the same temperature $T$, which leads to a modification in the structure of the $P - v$ diagram.
\subsection{Joule-Thompson expansion}
To learn more about the proposed thermodynamics of black holes, we approach the Joule-Thomson expansion \cite{2000,20012}. Keeping the charge fixed, the Joule-Thomson coefficient can be written as follows
\begin{equation}
\mu=\left( \dfrac{\partial T}{\partial P} \right)_{M}=\dfrac{1}{C_{P}} \left[ T \left( \dfrac{\partial V}{\partial T} \right)_{P}-V  \right]. \label{m}
\end{equation}
For the sake of future comparison, the equation of state for such a black hole could be expressed in terms of thermodynamic volume. Considering equations (\ref{V}), (\ref{P}) and (\ref{xxx}), we can obtain the temperature as a function of volume and pressure
{\footnotesize
\begin{equation}
T =\frac{  8\pi P (3+2A)(1+2A)\left( \dfrac{3V}{4\pi}\right) ^{\dfrac{4}{3+2A}}+3(1+B)(1+2A)\left( \dfrac{3V}{4\pi}\right) ^{\dfrac{2}{3+2A}}    -3 Q^{2} \! \left(2 A^{3}-5 A^{2}+1\right)  }{12\pi(1+2A)\left( \dfrac{3V}{4\pi}\right) ^{\dfrac{3}{3+2A}}}. \label{t}
\end{equation}}
  Using equation (\ref{t}) and the second part of equation (\ref{m}), we can get  the temperature associated with a zero Joule-Thomson coefficient. In fact, the repeated inversion temperature $T_i$ can be found to be
{\footnotesize \begin{equation}
T_i=\frac{  8\pi P\left(1+2A\right) \left( 3+ 2A \right)\left( \dfrac{3V}{4\pi}\right) ^{\dfrac{4}{3+2A}}-3(1+B)(1+2A)\left( \dfrac{3V}{4\pi}\right) ^{\dfrac{2}{3+2A}}   -3 Q^{2} \! \left(2 A^{3}-5 A^{2}+1\right) }{12\pi \left(1+2A\right) \left( 3+ 2A \right)\left( \dfrac{3V}{4\pi}\right) ^{\dfrac{3}{3+2A}}}.
\end{equation}}
After certain computations,  this temperature  can be shown to be
\begin{equation}
T_{i}=\frac{8 \pi P_{i}  \left(1+2A\right) \left( 3+ 2A \right)  r_h^{4}-3 \left(1+2 A \right)\left(1+B \right) r_h^{2}+9 Q^{2} \! \left(2 A^{3}-5 A^{2}+1\right)}{12  \pi  \left(1+2A\right) \left( 3+ 2A \right)  \,r_h^{3}}.
\label{ti1}
\end{equation}
 where $P_{i}$  is the inversion pressure. Using  Eq. (\ref{t}),   we  obtain
\begin{equation}
T =\frac{8 \pi P  \left(1+2A\right) \left( 3+ 2A \right) r_h^{4}+3 \left(1+2 A \right)\left(1+B \right) r_h^{2}-3 Q^{2} \! \left(2 A^{3}-5 A^{2}+1\right)}{12 \pi  \left(1+2 A \right)  \,r_h^{3}}.\label{ti2}
\end{equation}
Subtracting Eq. (\ref{ti1}) form Eq. (\ref{ti2}),  we  get  the algebraic equation
 \begin{equation}
8 \pi  P_{i}C   \,r_{h}^{4}-6 D  Q^{2}+3 K   r_h^{2}=0,
 \end{equation}
where   one  has used \begin{equation}
\begin{aligned}
D=&\dfrac{\left(9+A\right)}{\left( 1+2A\right)} \left(2A^3-5A^2+1 \right) \\
C=&\left( 3+2A\right)  \left( 1+A\right)\\
K=& \left( 1+B\right) \left( 2+A\right),
\end{aligned} \end{equation}
  By handling  this equation, we can obtain  four roots. However, only one of them has physical significance, while  the others are either complex or negative, which  must be highlighted.  We are only interested  in the real and positive root given by \begin{equation}
r^{i}_h = \frac{\sqrt{  \sqrt{32 \pi  P_{i} C D Q^{2}+9 K^{2}}-3 K}}{4 \sqrt{\pi  P_{i} C}}.
\end{equation}
At zero inversion pressure $P_i=0$, the inversion temperature takes a  minimum value
 \begin{equation}
T_{i}^{min}=\dfrac{(1+B)\sqrt{1+B}}{4\pi Q (3+A)\sqrt{(A+3)(2A-1)(A(A-2)-1)}}. 
 \end{equation}
This generates  a  ratio between minimum inversion and critical temperatures  expressed as follows
 \begin{equation}
\zeta= \dfrac{T_{i}^{min}}{T_{c}}=\dfrac{3 \sqrt{6}       \sqrt{2+A}  }{4(A+3) \sqrt{A+3}}.
 \end{equation}
Considering small values of $A$,   we get 
 \begin{equation}
\zeta=  \frac{1}{2}-\frac{A}{8}+\frac{5 A^{2}}{192}+O\! \left(A^{3}\right).
 \end{equation}
Taking  $A=0$, we recover  the usual result of charged solutions  $\zeta= \dfrac{1}{2}$ reported  in \cite{2000,2001,2002,20012,20013}.  This shows that the result obtained is perfectly consistent with the universal behavior of charged AdS black holes with respect to charge $Q$.
\subsection{Phase transitions}
\begin{figure}[h!]
    \centering
    \begin{tabular}{cc}
       \includegraphics[width=7cm,height=7cm]{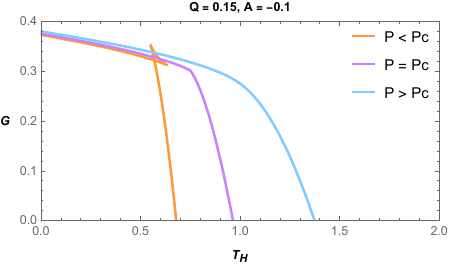} & 
         \includegraphics[width=7cm,height=7cm]{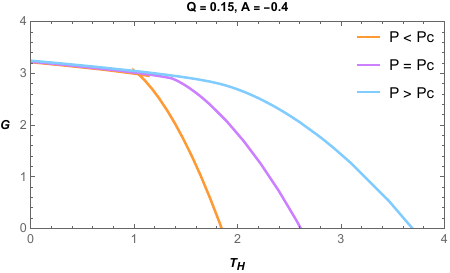} 
    \end{tabular}
    \begin{tabular}{cc}
       \includegraphics[width=7cm,height=7cm]{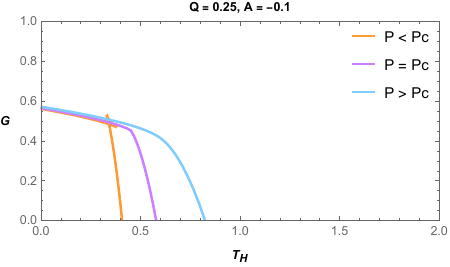} & 
         \includegraphics[width=7cm,height=7cm]{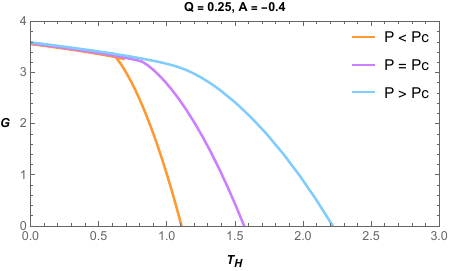} 
    \end{tabular}
    \caption{Gibbs free energy in terms of the temperature for different values of $P$ and $Q$, with $B=1$.}
    \label{7}
\end{figure}

To analyze the  phase transitions, we evaluate  the Gibbs free energy using the following relationship  
\begin{equation}
   G = M - T S.
\end{equation}
This quantity is found to be 
\begin{equation}
G =\frac{-8 \pi  P \left(1+2 A \right) r_h^{4}+3( B +1)r_h^2-9 Q^{2} \! \left(A \! \left(A -2\right)-1\right)}{12 \left(1+2 A \right) \! \left(1+A \right) r_h^{1-2 A}}.
\end{equation}
Taking $A=B=0$, we recover the Gibbs free energy of the  ordinary charged black hole 
\begin{equation}
G =\frac{-8 \pi  P  r_h^{4}+3r_h^2+9 Q^{2}}{12 r_h}
\end{equation}
reported in \cite{17}.  Using the critical thermodynamical quantities, the $G-T_H$ curves are presented  in Fig(\ref{7}).
It should be noted that the $G$–$T_H$ curves,   describing  the Gibbs free energy as a function of temperature, display similar qualitative behaviors for different values of the critical pressure $P_c$. More specifically, for the  pressures below the critical value ($P < P_c$), the characteristic swallow-tail  structure appears in the $G$–$T_H$ diagram. This characteristic is typical of first-order phase transitions, which mark a transition between the small and large black hole phases. Furthermore, increasing the values of the electric charge $Q$ or modifying the parameter
$A$ controls the position and the  shape of the curves, thus the temperature and Gibbs free energy values at phase transitions. This thermodynamics is extremely close to that of Van der Waals fluids, once more supporting the analogy between black hole systems and classical fluid models.

\section{Conclusions}
In this work, we have presented a new class of charged black holes by introducing Dunkl derivatives in the  four-dimensional spacetime. To construct these solutions, we have first computed the Ricci tensor and Ricci scalar using the Christoffel symbols, and then substituted them into the modified Einstein field equations.  These computations have provided a differential equation solved by the black hole metric function of charged Dunkl black holes. After that,  we have subsequently investigated the effect of the  charge $Q$  on the thermodynamical properties by calculating the associated quantities. To analyze  the thermal stability, we have determined the heat capacity. Furthermore, we have explored the $P$–$v$ criticality behavior by computing the critical pressure $P_c$, the critical temperature $T_c$, and the critical specific volume $v_c$ in terms of the charge $Q$ and two parameters, $A$ and $B$, which encode information about the Dunkl reflections. Notably, we have shown that the ratio $\frac{P_{c}v_{c}}{T_{c}}$ is a universal quantity with respect to the charge $Q$ and the parameter $B$. By taking the limit $A = 0$, we have recovered the behavior of a Van der Waals fluid. Regarding the Joule–Thomson expansion, we have revealed both similarities and differences compared to Van der Waals fluids. Finally, we have examined the phase transitions by computing the Gibbs free energy.

This work has left certain open questions. A natural direction for a  future study is to consider the inclusion of additional internal and external parameters, such as the spin parameter producing  rotating black holes.  It would be possible to use CUDA codes exploited in machine learning methods to impose additional constraints on the black hole parameter based on new insights  in the thermodynamic context.\\

  {\bf Data availability}\\
  Data sharing is not applicable to this article.

\section*{Acknowledgements}
The author  would like to thank  A. Belhaj  for indispensable discussions,  comments, and  encouragement.  She would also like to thank  S. E. Baddis, H. Belmahi  for  scientific help and collaborations. She would also  thank the anonymous referees for their helpful comments, suggestions, and remarks that improved the quality of the manuscript.   This work was done with the support of the CNRST in the frame of the PhD Associate Scholarship Program PASS.

\end{document}